\documentclass[aps,11pt,amsmath,amssymb]{revtex4}
\usepackage{dcolumn}
\usepackage{bm}
\begin{document}
\title{Tunneling through the quantum horizon}
\author{Michele Arzano}
\email{arzano@physics.unc.edu}
\affiliation{Institute of Field Physics\\Department of Physics and Astronomy\\
University of North Carolina\\
Chapel Hill, North Carolina 27599-3255, USA}
\begin{abstract}
\begin{center}
{\bf Abstract}
\end{center}
The emergence of quantum-gravity induced corrective terms for the probability of emission of a particle from a black hole in the Parikh-Wilczek tunneling framework is studied. It is shown, in particular, how corrections might arise from modifications of the surface gravity due to near horizon Planck-scale effects.  Our derivation provides an example of the possible linking between Planck-scale departures from Lorentz invariance and the appearance of higher order quantum gravity corrections in the black-hole entropy-area relation.  
\end{abstract}
\maketitle
\section{Introduction}
Classical black holes are perfect absorbers: they accrete their (irreducible) mass and no fraction of it can escape as there are no classical allowed trajectories crossing the horizon on the way out.   This particular behavior suggests an interpretation of the black hole area as an entropy-like quantity.  More than thirty years ago Bekenstein \cite{Bekenstein:1971hc}, with an elegant argument, showed that the entropy of a black hole must, in fact, be proportional to its surface area measured in units of Planck length squared.  The key point of his derivation was  the inclusion of the quantum mechanical properties of a particle crossing the black-hole horizon.  For this reason the entropy-area  relation can be viewed as a first step towards the understanding of  ``quantum" properties of black holes.  Hawking's discovery 
\cite{Hawking:1974sw} that quantum fields on a Schwarzschild background  do indeed predict a thermal flux of particles {\it away} from the horizon confirmed that the black hole entropy/area is in all senses a thermodynamic quantity and it is legitimate to define a temperature that corresponds to a ``physical" temperature associated with the radiation.\\
It is interesting to note how the inclusion of quantum effects allows, for particles in a Schwarzschild geometry, to propagate through classically forbidden regions.
This suggests that it should be possible to describe the black hole emission process, in a semiclassical fashion, as quantum tunneling.
Parikh and Wilczek \cite{Parikh:1999mf} (see also \cite{Parikh:2004ih, Parikh:2004rh})  showed how such a description of black hole radiance is possible if one considers the emission as a transition between states with the same energy.  In this way the lowering of the mass of the black hole during the process and the related change in the radius set the barrier through which the particle tunnels.  The resulting probability of emission differs from the standard Boltzmann factor by a corrective term which depends on the ratio of the energy of the emitted particle and the mass of the hole.  The appearance of the correction causes the emission spectrum to be non-thermal. This reflects the fact that in order to describe transitions in which the energy of the emitted particle-black hole system does not change one must take into account  the particle's self-gravitation.  In the limit when the energy of the emitted particle is small compared with the mass of the  black hole the emission spectrum becomes thermal and  Hawking's result is recovered.\\
In the tunneling framework the  Bekenstein-Hawking entropy-area relation can be deduced from the form of the emission probability.  In fact the latter is proportional \cite{Parikh:2004ih, Parikh:2004rh} to a phase space factor depending on the initial and final entropy of the system which multiplies the square of the quantum mechanical amplitude for the process.  A black hole entropy given by the Bekenstein-Hawking formula $S_{BH}=\frac{A}{4}=4 \pi M^2$ corresponds to the Parikh-Wilczek result for the tunneling amplitude. \\
In this letter we ask how the Parikh-Wilczek tunneling picture might be affected by the presence of Planck-scale effects for the near-horizon emission process.  
In particular we consider the scenario in which quantum gravity departures from Lorentz symmetry affect a particle's special relativistic energy-momentum dispersion relation.
In \cite{Amelino-Camelia:2004xx} using the Bekenstein argument ``in reverse", it was shown how, in loop quantum gravity, a deformed energy-momentum dispersion relation can be related to the appearance of a logarithmic correction to the semiclassical entropy-area law.
Besides loop quantum gravity \cite{Rovelli:1996dv, Ashtekar:1997yu,Kaul:2000kf}, a logarithmic correction to entropy/area law has also emerged in string theory \cite{Strominger:1996sh, Solodukhin:1997yy} and other approaches \cite{Fursaev:1994te} (see also \cite{Page:2004} and references therein).\\
Here we show how, in the spirit of  \cite{Amelino-Camelia:2004xx}, the log-corrected entropy-area relation naturally emerges in the tunneling picture when Planck-scale effects are taken into account.
We start with a brief review of the standard tunneling argument and then proceed to modify it 
with the inclusion of near horizon Planck-scale effects.

\section{Tunneling through the horizon}
We obtain here an expression for the tunneling probability of a spherical shell through the horizon of a Schwarzschild black hole.  The two main ingredients of \cite{Parikh:1999mf} are the use of the WKB approximation for the tunneling probability and an effective action describing the system which includes the shell's self-gravitation.  The first approximation is valid since wave packets propagating from near the horizon are arbitrarily blue-shifted there, the geometrical optics limit applies and we can treat the shell as a particle.
In the WKB approximation the tunneling probability is a function of the imaginary part of the action
\begin{equation}
\label{tunnelampl}
\Gamma\sim e^{-2\,\mathrm{Im}\, S}\,\, .
\end{equation}
The action needed to compute the emission probability can be found in \cite{Kraus:1994by}.  There the corrections to the geodesic motion of a spherical shell due to self-gravitation in a Schwarzschild geometry were calculated and their consequences for the Hawking radiation spectrum were studied 
(see also \cite{Keski-Vakkuri:1996xp}).
One starts by considering the metric for a general spherically symmetric system in ADM form 
\begin{equation}
ds^2=-N_t(t,r)^2dt^2+L(t,r)^2[dr+N_r(t,r)dt]^2+R(t,r)^2d\Omega^2\,  .
\label{metricADM}
\end{equation}
Once the action for the hole-shell system has been written in Hamiltonian form, the dependence from all the momenta, but the one conjugate to the shell radius, can be eliminated using the constraints of the theory.  Integrating over the gravitational degrees of freedom and fixing the gauge appropriately 
($L=1$ $R=r$) \footnote{This choice of the gauge corresponds to a
particular set of coordinates for the line element (Painleve' coordinates) which is particularly useful to study across horizon phenomena being non-singular at the horizon and having Euclidean constant time slices (for more details see \cite{Kraus:1994fh}).} one obtains the following effective action for a massless self-gravitating spherical shell 
\begin{equation}
S=\int dt \left(p_c\dot{\hat{r}}-M_+\right)\, .
\end{equation}
Here $p_c$ is the momentum canonically conjugate to the radial position of the shell, and
$M_+$ is the total mass of the shell-hole system which plays the role of the Hamiltonian.
In terms of the black hole mass $M$ and the shell energy $E$ we have $M_+=M+E$.
Details of the lengthy derivation can be found in \cite{Kraus:1994by}.
The trajectories which extremize this action are the null geodesics of the metric
\begin{equation}
ds^2=-[N_t(r; M+E)dt]^2+[dr+N_r(r; M+E)dt]^2+r^2d\Omega^2\,  ,
\label{linel}
\end{equation}
for which
\begin{equation}
\frac{dr}{dt}= N_t(r; M+E)-N_r(r; M+E)\, .
\label{geodesic1}
\end{equation}
An explicit form for the line element (\ref{linel}) can be obtained from the expressions of $N_t$ and $N_r$ given by the constraint equations \cite{Kraus:1994by}
\begin{equation}
N_t=\pm1\,\,\,;\,\,\,  N_r=\pm\sqrt{\frac{2M_+}{r}}\,\,.
\end{equation}
In \cite{Parikh:1999mf} 
the total mass of the system is kept fixed while the hole mass is allowed to vary. This means that 
the mass parameter $M_+$ is now $M_+=M-E$.
One then has the following expression for a radial null geodesic
\begin{equation}
\dot{r}=\pm1-\sqrt{\frac{2(M-E)}{r}}\, .
\label{geodesic2}
\end{equation}
Now consider the emission of a spherical shell for which
\begin{equation}
\label{shellaction}
\mathrm{Im}\, S=\mathrm{Im}\int_{r_{in}}^{r_{fin}}p_r dr\, .
\end{equation}
$r_{in}$ and $r_{fin}$ are radial positions just inside and outside the barrier through which the particle is tunneling. 
To calculate $\mathrm{Im}\, S$ we can use Hamilton's equation, $\dot{r}=\frac{\partial H}{\partial p}$,
\begin{equation}
\mathrm{Im}\, S=\mathrm{Im}\int_{r_{in}}^{r_{fin}}p_r dr=
\mathrm{Im}\int_{r_{in}}^{r_{fin}}\int_M^{M-E}\frac{dH'}{\dot{r}} dr\, .
\end{equation}
The Hamiltonian is $H'=M-E'$, so the imaginary part of the action reads
\begin{equation}
\mathrm{Im}\, S=-\mathrm{Im}\int_{r_{in}}^{r_{fin}}\int_0^E\frac{dE'}{\dot{r}} dr\, .
\label{imsf}
\end{equation}
Using (\ref{geodesic2}) and integrating first over $r$ one easily obtains
\begin{equation}
\Gamma\sim \exp\left(-8\pi M E\left(1-\frac{E}{2M}\right)\right)\, ,
\label{proba}
\end{equation}
which, provided the usual Bekenstein-Hawking formula $S_{BH}=A/4=4\pi M^2$ is valid, corresponds to 
\begin{equation}
\label{Gammaf}
\Gamma\sim \exp\left[S_{BH}(M-E)-S_{BH}(M)\right]\, .
\end{equation}
If one integrates (\ref{imsf}) first over the energies it is easily seen that in order
to get (\ref{proba}) we must have $r_{in}=M$ and $r_{out}=M-E$.  So according to what one would expect from energy conservation, the tunneling barrier is set by the shrinking of the black hole horizon with a change in the radius related to the energy of the emitted particle itself.\\
\section{A tunnel through the quantum horizon}
The probability of emission of a shell with energy $E$, in the presence of back-reaction effects, put in the form (\ref{Gammaf}) is highly suggestive.
It is what one would expect from a quantum mechanical
calculation of a transition rate where, up to a factor containing the square of the amplitude of the process,
\begin{equation}
\label{ampliG}
\Gamma\sim \frac{e^{S_{fin}}}{e^{S_{in}}}=\exp\left(\Delta S\right)\, .
\end{equation}
In other words the emission probability is proportional to a phase space factor which depends on the initial and final entropy of the system.  The entropy is directly related to the number of micro-states available to the system itself.\\
This observation calls for an immediate generalization.  As we stressed in the Introduction,  derivations of the black hole entropy-area relation in several quantum gravity scenarios besides reproducing the familiar Bekenstein-Hawking linear relation give a leading order correction with a lo\-garithmic \footnote{A similar logarithmic term has also emerged in the calculation of 
one-loop quantum corrections to the entropy-area law in ordinary QFT \cite{Fursaev:1994te}. } 
dependence on the area \footnote{We now switch from $k=\hbar=c=G=1$ units of the previous sections to $k=\hbar=c=1$ to keep track of the Planck-scale suppressed terms.}
\begin{equation}
\label{logcorrs}
S_{QG}=\frac{A}{4L_p^2}+\alpha \ln \frac{A}{L_p^2}+O\left(\frac{L_p^2}{A}\right)\,\, .
\end{equation}
Now consider the emission of a particle of energy $E$ from the black hole.  One might expect that a derivation of the emission probability in a quantum gravity framework in presence of back-reaction would lead to an expression analogous to (\ref{ampliG}) with the usual Bekenstein-Hawking entropy $S_{BH}=\frac{A}{4L_p^2}$ replaced by (\ref{logcorrs}) i.e.
\begin{equation}
\Gamma\sim \exp{(S_{QG}(M-E)-S_{QG}(M))}\, .
\end{equation}
The previous expression written in explicit form reads
\begin{equation}
\Gamma(E)\sim \exp{(\Delta S_{QG})}=
\left(1-\frac{E}{M}\right)^{2\alpha}\exp \left(-8\pi GME\left(1-\frac{E}{2M}\right)\right)\, .
\label{prob}
\end{equation}
The exponential in this equation shows the same type of non-thermal deviation found in \cite{Parikh:1999mf}.  In this case, however, an additional factor depending on the ratio of the energy of the emitted quantum and the mass of the black hole is present.  A discussion of the possible consequences of the additional factor for the fate of the black hole in its late stages of evaporation and the appearance of statistical correlation between quanta emitted with different energies can be found in \cite{ArzMedVag}.\\
 In \cite{Amelino-Camelia:2004xx} the present author, Amelino-Camelia and Procaccini showed how in the context of loop quantum gravity a logarithmic corrective term to the entropy-area law of the type present in (\ref{logcorrs}) can be related to a modification of the energy-momentum dispersion
relation for a massless particle propagating in flat space-time
\begin{equation}
\label{modisprel}
p^2\simeq \left(1+\eta L_p^2 E^2\right)E^2\,\, .
\end{equation}
with $\eta=\frac{2\pi}{3}\alpha$ \footnote{In loop quantum gravity $\alpha$ is a negative coefficient whose exact value was once an object of debate (see e.g. \cite{Ghosh:2004rq})
but has since been rigorously fixed at $\alpha=-1/2$ \cite{Meissner:2004}.}.\\
As we already observed in the previous sections, the black hole radiation spectrum seen from an observer at infinity is dominated by modes that propagate from ``near" the horizon where they have arbitrarily high frequencies and their wavelengths can easily go below the Planck length \cite{Jacobson:1991gr,Jacobson:1993hn}.  It turns out then that a key assumption in all the derivations of the Hawking radiation is that the quantum state near the horizon looks, to a freely falling observer, like the Minkowski vacuum.  In other words Lorentz symmetry should hold up to extremely short scales or very large boosts.
It is plausible then that the motion of our particle tunneling through the horizon might be affected by Planck scale corrections of the type (\ref{modisprel}).  These type of modified dispersion relations
have, in fact, been proposed as low-energy quantum gravity effects, which deform
\cite{Amelino-Camelia:2000ge, Amelino-Camelia:2000mn} or break \cite{Amelino-Camelia:1997gz, Gambini:1998it} 
Lorentz symmetry (see also \cite{Ng:2003jk}).\\ 
One would expect that an analysis analogous to the ones of the previous sections with opportune
modifications should lead to
a result of the form (\ref{Gammaf}) with $S_{BH}$ replaced by $S_{QG}$. We verify here the above conjecture with an explicit example. As seen in the previous section the tunneling amplitude (\ref{tunnelampl}) for the emission of a spherical shell can be derived from 
\begin{equation}
\mathrm{Im}\, S=\mathrm{Im}\int_{r_{in}}^{r_{fin}}p_r dr=
\mathrm{Im}\int_{r_{in}}^{r_{fin}}\int_0^H\frac{dH'}{\dot{r}} dr=
-\mathrm{Im}\int_{r_{in}}^{r_{fin}}\int_0^{E}\frac{dE'}{\dot{r}} dr
\,\, .
\end{equation}
Now we proceed to evaluate the integral without using an explicit form for
the null geodesic of the spherical shell in terms of its energy.
In fact, near the horizon, where our integral is being evaluated, one has
\begin{equation}
N_t(r; M)-N_r(r; M)\simeq (r-R)\,\kappa(M)+O((r-R)^2)
\end{equation}
where R is the Schwarzschild radius and $\kappa(M)$ is the horizon surface gravity. 
Taking into account self-gravitation effects, $\dot{r}$ can be approximated by
\begin{equation}
\dot{r}\simeq (r-R)\,\kappa(M-E)+O((r-R)^2)\, .
\end{equation}
We can then write
\begin{equation}
\mathrm{Im}\, S=-\mathrm{Im}\int_{r_{in}}^{r_{fin}}\int_0^{E}\frac{dE'}{(r-R)\,\kappa(M-E')} dr
\,\, .
\end{equation}
Integrating over $r$, using the Feynman prescription\footnote{The pole is
moved in the lower half plane as in \cite{Parikh:1999mf}.} for the pole on the real axis $r=R$, we get
\begin{equation}
\mathrm{Im}\, S=-\pi\int_0^{E}\frac{dE'}{\kappa(M-E')}
\,\, .
\label{Ims}
\end{equation}
The surface gravity appearing in the above integral carries quantum gravity
corrections coming from Planck-scale modifications of near horizon physics related to (\ref{modisprel}).  
These modifications are such that they reproduce via the first law of black hole thermodynamics, $dE'=dM'=\frac{\kappa(M)}{2\pi}dS$, the quantum gravity corrected entropy-area law (\ref{logcorrs}) 
\cite{Amelino-Camelia:2004xx}.
Using the first law, (\ref{Ims}) becomes
\begin{equation}
\mathrm{Im}\, S=-\frac{1}{2}\int_{S_{QG}(M)}^{S_{QG}(M-E)}dS=\frac{1}{2}[S_{QG}(M)-S_{QG}(M-E)]
\end{equation}
which leads to a probability of emission
\begin{equation}\label{prob2}
\Gamma(E)\sim \exp{(-2\mathrm{Im}S)}=
\left(1-\frac{E}{M}\right)^{2\alpha}\exp \left(-8\pi GME\left(1-\frac{E}{2M}\right)\right)\, 
\end{equation}
analogous to (\ref{prob}).\\
To summarize: we showed how Planck-scale effects can be incorporated in the Parikh-Wilczek tunneling picture.  We were able to obtain a form of the emission probability which includes both back-reaction and quantum gravity effects. Our derivation gives an idea of how near horizon physics provides an excellent arena for studying the interplay of seemingly different aspects of quantum gravity as the number of microscopic degrees of freedom of a black hole and possible Planck-scale modifications of space-time symmetries.\\
\begin{center}
{\bf Acknowledgements}
\end{center}
I would like to thank Giovanni Amelino-Camelia for discussions and valuable suggestions and Jack Ng for useful comments. I also thank the Department of Physics of the University of Rome "La Sapienza" for hospitality during December 2004.

\end{document}